\documentclass[envcountsames]{llncs}
\usepackage{graphicx}
\usepackage{comment}
\usepackage{url}
\begin{document}

\title{A Stateless Transparent Voting Machine}

%\email{webmaster@marysville-ohio.com}
%\affiliation{%
%  \institution{Institute for Clarity in Documentation}
%  \streetaddress{P.O. Box 1212}
%  \country{USA}
%\author{Lars Th{\o}rv{\"a}ld}
%\affiliation{%
%  \institution{The Th{\o}rv{\"a}ld Group}
%  \streetaddress{1 Th{\o}rv{\"a}ld Circle}
%\email{larst@affiliation.org}

%\author{anonymous}
\author{Juan E. Gilbert, Jean D. Louis }

\authorrunning{Gilbert & Louis}
\institute{University of Florida, Gainesville FL 32611, USA}
%\and Springer Heidelberg,


\maketitle
\begin{abstract}
Transparency and security are essential in our voting system, and voting machines. This paper describes an implementation of a stateless, transparent voting machine (STVM). The STVM is a ballot marking device (BMD) that uses a transparent, interactive printing interface where voters can verify their paper ballots as they fill out the ballot. The transparent interface turns the paper ballot into an interactive interface. In this architecture, stateless describes the machine's boot sequence, where no information is stored or passed forward between reboots. The machine does not have a hard drive. Instead, it boots and runs from read-only media. This STVM design utilizes a Blu-ray Disc ROM (BD-R) to boot the voting software. This system's statelessness and the transparent interactive printing interface make this design the most secure BMD for voting. Unlike other voting methods, this system incorporates high usability, accessibility, and security for all voters. The STVM uses an open-source voting system that has a universally designed interface, making the system accessible for all voters independent of their ability or disability. This system can make voting safer by simultaneously addressing the issue of voters noticing a vote flip and making it difficult for a hack to persist or go unmitigated.
\end{abstract}




\section{Introduction}
Ballot marking devices (BMD) are systems that will produce physical ballots after voters select their options on a computing device using a touchscreen and/or accessibility interfaces/devices. BMD have advantages for making voting accessible and easy to use. Voting is an essential part of how Democracy works. BMD enable more people to participate in the voting process. BMD were used in 47 of the 50 states in the 2020 U.S. general elections \cite{ballotpedia}. However, recently, BMD have been criticized for voter verification [5]. Voter verification refers to voters reviewing their ballots for accuracy before their ballot is officially cast. For example, if several BMD become compromised and many voters fail to verify their printed ballots, the outcome of an election could be altered and go undetected. Rivest and Wack introduced the notion of software independence: ``A voting system is software-independent if an undetected change or error in its software cannot cause an undetectable change or error in an election outcome \cite{rivest2008notion}.” BMD are mostly software-independent, specifically when voters verify their printed ballots. However, if voters cannot or do not verify their printed ballot, then this can compromise software independence. Recent studies have suggested that an insufficient number of voters actually verify the printed ballot \cite{bernhard2020can,kortum2021voter}. In response to these studies, a transparent, interactive printing interface, which we will refer to as the transparent voting machine, was created for use in elections \cite{gilbert2021study}.  
\begin{figure*}
    \begin{center}
        \includegraphics[scale = 0.80]{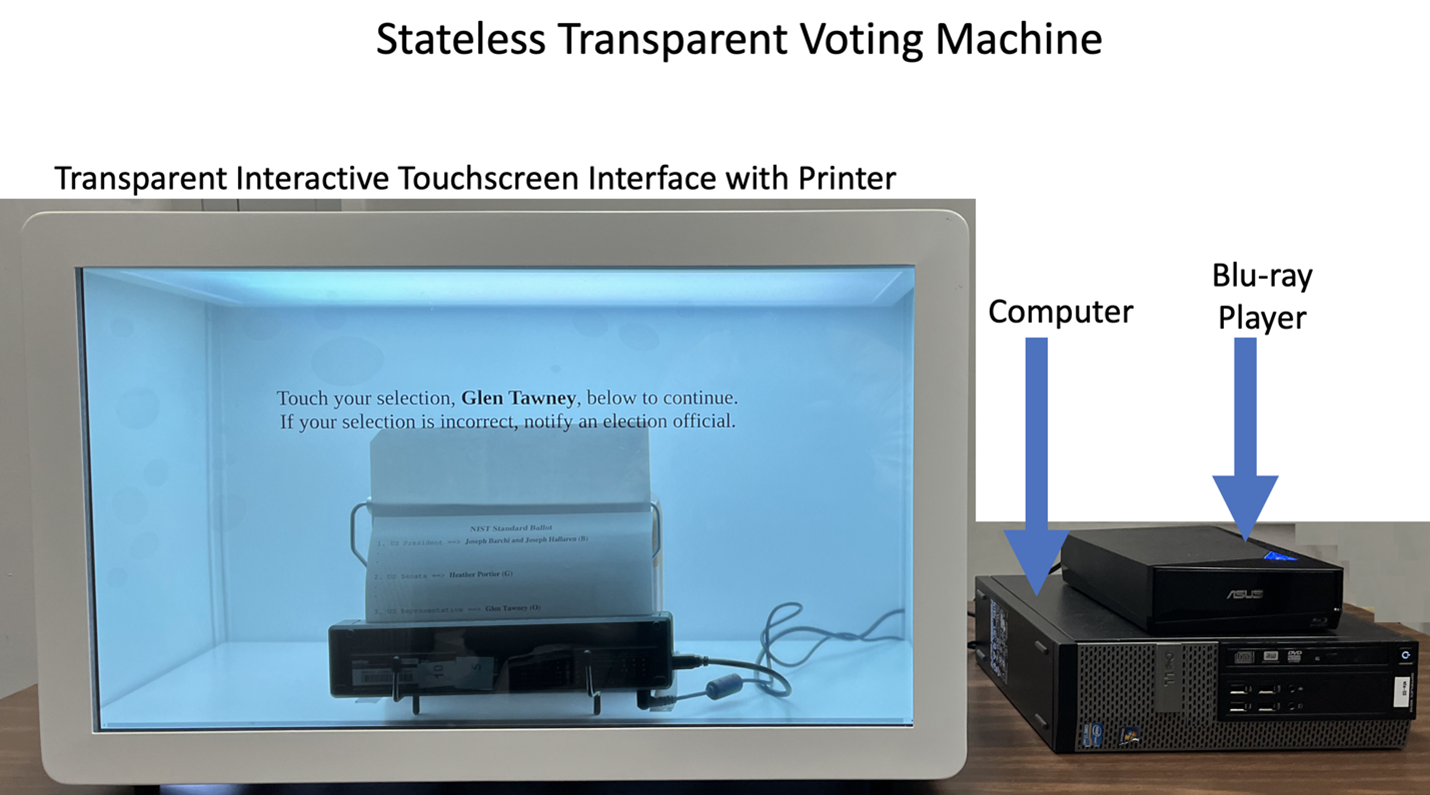}
        \caption{Stateless Transparent Voting Machine Prototype}
        \label{TVI}
    \end{center}
\end{figure*}

\section{Transparent Voting Machine}
The transparent voting machine (TVM) implements an interactive printing interface, see Figure \ref{TVI}. The interactive printing interface has a transparent screen in front of the printer, which holds a blank paper ballot like other BMD, but the voter interacts with the transparent touchscreen. Before the voter can move to another contest, the voter must print their selection and verify the selection. As voters select their options on the transparent touchscreen, they can witness their selection being printed behind the glass screen. After each selection is printed, the voter must touch the screen to confirm that the printed ballot is what they selected on the screen. This confirmation step is required to advance in the voting process. Unlike previous BMD, TVM has a voter verification step built in. Studies conducted with the TVM suggest that for an election using a transparent voting machine, it is unlikely that an election could be hacked such that the tampering would go unnoticed. TVM has also been proven to be usable \cite{gilbert2021study}. In In Bernhard et al. \cite{bernhard2020can}, only 40\% of voters reviewed their ballots. Kortum et al. \cite{kortum2021voter}, only 17.6\% of error detection from voters. The TVM had a 77\% detection rate \cite{gilbert2021study}. The TVM secures the paper ballot through voter verification. Prior to voter verification, it is crucial to ensure the voting machine has not been compromised. Even with the TVM, where voters will notice any incorrect marks on their ballot, there is still a need to limit the infusion of malware or hardware that could tamper with the voting machine.

\section{Stateless Transparent Voting Machine}
The stateless transparent voting machine (STVM) uses the transparent interactive printing interface from the transparent voting machine described in the previous section but adds a new component: a stateless computer. The computer does not have a hard drive or any network connection; therefore, the machine has no memory or the ability to store anything that will persist after a hard reboot, hence the term stateless. The voting machine also has a Blu-ray drive and a connected printer. The voting machine is powered by a read-only Blu-ray Disc ROM (BD-R), storing the entire operating system with voting software. Lastly, the voting machine is connected to the transparent touchscreen, as seen in Figure \ref{TVI}. In this prototype, the computer is a standard desktop with network cards and hard drives removed. The operating system is a stripped-down Ubuntu Linux distribution, and the system image is burned to a BD-R. Blu-ray Discs are used over regular DVD drives because they improve boot-up speed. The printer is a standard small form factor thermal printer, set to print in continuous mode. There are several advantages to the STVM design. The following sections will describe these benefits by comparing the STVM against current BMD and hand-marked paper ballots.

\section{BMD vs. STVM}
BMD have several advantages that are shared with the STVM. These include accessibility, usability, and security features. With respect to accessibility, they both provide features such as audio, accessible switches, interface resizing, etc., that enable voters to vote privately and independently. The STVM runs an open-source software called Prime III \cite{smarr2017prime} that has demonstrated and has well-received accessibility features. Furthermore, the Prime III voting software was certified for statewide use in New Hampshire as their accessible voting machine \cite{brooks_2018,exeter_new_hampshire}. In addition, Butler County, Ohio, uses Prime III as its remotely accessible voting software \cite{verified_voting}. With respect to accessibility, BMD and STVM share common features, and both provide voters with privacy and independence. 

With respect to usability, both BMD and the STVM employ touchscreens that provide easy-to-navigate interfaces. Users who are familiar with touchscreens on computers, tablets, and phones are likely to see voting on a touchscreen as a natural transition \cite{gilbert2021study}. Therefore, BMD and the STVM are both usable. However, it is possible to deploy a poorly designed user interface on a BMD that would compromise the system's usability, accessibility and security..  

\subsection{Security Features}
Modermn BMD are not stateless machines. They have hard drives where the operating system, voting system, and other components reside. As such, it has been suggested that a standard BMD can be infected with malware that misbehaves during the election and disappears once the election is over \cite{liu2011behavior,feldman2006security}. With the STVM, this does not happen. There is no hard drive to inject malware. The only place to inject malware would be the BD-R or the BIOS. If malware is inserted into the BD-R, it can certainly be discovered because of its read-only nature. Blu-ray Discs are burned using a laser that encodes the target data onto the disk's surface, known as optical disk recording. The malware cannot erase itself. Once installed, it lives on the BD-R forever. In an attack scenario where the attacker has access to the BD-R and attempts to swap or modify it to flip votes, we know it is unlikely that tampering would go unnoticed, and the discrepancies in the disk are easily discoverable. With respect to the BIOS attack, two main things are needed: extensive access to the voting machines and specialized expertise. The STVM setup protocol recommends installing the BIOS from a read-only DVD-ROM before the election begins. Therefore, if malware were inserted into the BIOS before the election, it would be overwritten prior to starting the election. Even if malware made its way onto the BD-R, Gilbert et al. \cite{gilbert2021study} suggest that with the transparent interface, voters would notice their votes being flipped and could alert election officials so that an investigation could begin. The BD-R's checksum can be verified on a separate machine to validate the BD-R. 

In Sastry et al. \cite{sastry2006designing}, the authors present four design principles for voting machines that this STVM inadvertently fulfills. Property 1, one voter's interaction with the machine should not affect another. The STVM is stateless and reboots between uses, so no data from one voter is present to another. Property 3, the system ``should be history independent and tamper evident," is addressed by using BD-R. Property 2: A ballot cannot be cast without the voter's consent to cast. Property 4 refers to the ballot not changing after selection. Properties 2 and 4 are satisfied by the STVM, which physically prints each user selection on a paper ballot and asks for confirmation to continue. 

\section{BMD Attacks and STVM Defense}
This design is resistant to many of the past attacks on BMD used in Feldman et al. \cite{feldman2006security} and Aviv et al. \cite{aviv2008security}. Many of these attacks are based on re-writable storage and BIOS tampering. Attempts to tamper with the screen calibration or physical components are detectable by voters and poll workers and are rectifiable. With these precautions, the STVM has an advantage over standard BMD. To clarify, this does not make BMDs insecure; these features just make the STVM more secure than standard BMD. The following section covers some popular attacks and STVM defenses.  

\subsection{Vote-Stealing Attacks}
In a vote-stealing attack, the goal is to transfer a vote cast for one candidate to another. In this scenario, the malicious code would keep the total number of votes the same to avoid raising an alarm. In the voting machine studied by Feldman et al.\cite{feldman2006security}, vote tally data was stored in flash memory and could be modified. In the proposed implementation of the attack, the malicious code would be inserted by modifying the binary of the source code or ``grafted into the operating system or bootloader". The attack would then be triggered during the active time of the election. The STVM has a defense to this attack class. For one, any modification to the source code while the machine runs does not persist. Also, the vote tally is not stored on the machine; this machine produces an auditable physical paper ballot.   
\subsection{Denial-of-Service Attacks}
Another attack class presented in \cite{feldman2006security} is the common Denial-of-Service(DOS). This attack aims to make voting machines unable to receive votes or be used during an election. To demonstrate this attack, the researchers created a user-space Windows executable to trigger a malicious bootloader to erase the contents of the inserted memory and onboard flash (which contains voting logs). Then, the computer is shut down \cite{feldman2006security}.
The stateless nature of the STVM means that there is nothing to delete. Also, user-space code modifications would not persist beyond a reboot. A simple defense to this attack is to reload the BIOS at the start of the election, overwriting potential boot loader malware and restoring the computer to a known state.
\subsection{Physical Security, Locks, and Seals Attacks} 
Physical security is an attack class discussed in \cite{aviv2008security}. In this attack class, attackers may attempt to gain unauthorized access to screen calibration or break part of the voting machine. The main attacker scenario we focus on in this paper is the attacker attempting to flip votes to change the election outcome. With this in mind, certain physical attacks remain out of scope; such attacks, include breaking the Blu-ray drive as a means of a DoS attack or other physical harm, like shattering the touch screen, are not the main focus. The downfall of these attacks is that they do not scale, cannot persist, and do not flip the users' votes undetected. 

\subsection{Remote-Control Attacks} 
The remote-control attack class is presented in \cite{abba2017security}. In this type of attack, the attacker would like to gain remote control of the system to either install malware, steal voting data, or impair the functions of the voting machines. To guard against this type of attack, the STVM does not have network drivers or non-volatile memory. An attacker would need to attach drives to every machine and re-establish connections on each reboot without voters noticing the vote being flipped. This attack does not scale and can easily be mitigated once discovered.
%https://ieeexplore.ieee.org/abstract/document/8252006
%election setup, voting, post-election

\subsection{Code Injection Attacks} 
Other attacks, such as ``Injecting Attacks," presented in Feldman et al.\cite{feldman2006security} and Aviv et al. \cite{aviv2008security} rely on the statefulness of the machine and do not present methods for flipping votes in a manner that would go undetected if applied to the STVM. This is also a difficult attack in practice because it must be done before the election. One problem with this is that the attacker cannot modify the attack based on real-time events such as the leading candidate or candidate dropout \cite{abba2017security}. With the STVM, users will notice injection attacks to flip votes. STVM has several mitigation strategies for this attack class. For one, Simple User-space malware injection will not persist beyond a reboot without non-volatile memory. Next, a new, clean BD-R could replace the current one. Finally, if necessary, Reloading the BIOS can quickly bring the machine back to a known state. If the malware was introduced by swapping the BD-R, the tamper-evident measure will be checked, and a byte-by-byte analysis will be performed to discover the discrepancies. The assumption made in this scenario is that there is a clean(unaffected) BD-R against which to compare. We also do not directly investigate attack scenarios related to social engineering or supply chain because our design does not have control of these steps, so these are out of scope.
%new clean ROM and BIOS, voter noticing

\subsection{DEF CON voting village}
This section discusses attacks performed on voting machines from DEF CON 25 \cite{blaze2017defcon} and 27 Voting Machine Hacking Village reports. Unlike the previous section, the DEF CON voting village was not part of a formal security study but was an opportunity for everyday hackers to test actual voting equipment. This section demonstrates the robustness of the STVM to previous attacks performed on voting machines. 

In the DEF CON 25 \cite{blaze2017defcon} report (held in 2017),  we will look at three main machines attacked. The AVS WinVote, which featured LAN communication, was compromised remotely due to IP address broadcasting. The attacker gained the ability to shut down the machine and change votes in the database remotely. Regarding physical attacks, the machine's USB slots were not protected. In the case of the AccuVote-TSx and ES\&S iVotronic systems, votes were recorded on internal flash, and hackers used physical tampering measures to gain access to this chip. These attacks would not work on the STVM. The STVM is an air gapped machine and does not have network drivers or network cards installed. Also, physical attacks on ports to upload code would not persist or flip votes undetected. The central defense against these vote tampering attacks is that the votes are not stored in the STVM's internal memory. The STVM produces paper ballots to be verified by the user and as part of a Risk-Limiting Audit \cite{stark2009risk} to provide statistical reassurance of election results.

In the DEF CON 27 report\footnote{https:\/\/harris.uchicago.edu\/files\/def\_con\_27\_voting\_village\_report.pdf} (held in 2019), the authors affirm the need for paper-backed ballots and Risk-Limiting Audits. This report also contained attacks that the STVM is robust to. At DEFCON 27, hackers highlighted the lack of tamper-evident seals and ``bloatware" still installed on some machines. For example, ES\&S: ExpressPoll Tablet Electronic Pollbook lacked physical protection and still had default Toshiba tablet apps like Netflix installed. Furthermore, the device still had wifi capabilities. The ES\&S Automark also lacked some tamper-evident seals but also featured an Ethernet port (after removing the case) and some unnecessary software elements. Participants were able to modify the firmware, crash the voting software, and modify configuration files. The STVM reduces the attack surface by removing unneeded apps from the operating system and would require tamper-evident seals widely used by other voting machines during elections. The successful attacks presented at DEF CON would not work on the STVM to cause the system to alter the election outcome through unnoticed vote flipping.

\section{Persistent Vote Flipping Hack Experiment}
This section discusses an experiment where a BMD was compared to the STVM against a malware hack. In this experiment, 2 Dell Optiplex computers were used. This experiment did not use any standard voting equipment, as voting machines are not easy to obtain for research experiments that involve malware. The Prime III open-source software \cite{smarr2017prime}, which has been used in local, state and federal elections in the U.S.A., was used as the voting system for this experiment. The first computer served as a control machine set up like a traditional BMD. The control machine had an attached printer and a hard drive. The experiment machine had an attached printer and a Blu-ray drive and did not have a hard drive. The experiment machine was set up as a STVM, as seen in Figure \ref{TVI}. A malware version of the Prime III software was created that would flip votes; specifically, the first contest on the ballot would be flipped. The goal was to see the effects of the malware before and after a reboot of the voting machine. 

The experiment began by installing Prime III on the control machine. Once Prime III was installed, votes were cast to ensure it was working correctly. This is a form of pre-election testing or logic \& accuracy testing \cite{hart-intercivic-nass-2022}. After voting on the control machine, the malware was inserted into the control machine using a USB disk and removed. This scenario mimics an insider attack where the malware is inserted on a machine before the election in secret. The control machine was rebooted and voting restarted. During voting, we observed the votes were being flipped; therefore, the malware attack was successfully installed and was available after rebooting the control machine. The malware software was identical to Prime III, except it flipped votes for the first contest on the ballot. The only way to notice the malware was that votes were being flipped on the paper ballot.   

The experiment machine was booted with Prime III on a read-only Blu-ray Disc ROM (BD-R). Once Prime III was loaded, votes were cast to confirm the system was functional as a form of pre-election testing as was done with the control machine. The same malware was then installed on the experiment machine to mimic an insider secretly infecting the machine before the election using the USB disk and the USB disk was removed. The experiment machine was rebooted and voting restarted. No votes where flipped. The malware was not able to persist on the machine. This is not a surprise given the stateless nature of the experiment machine. The next step involved installing the malware again during the voting session without rebooting the machine. Therefore, we inserted the USB disk, installed the malware, removed the USB disk and voted. During this test, the vote for the first contest was flipped. The malware was installed and working. The experiment machine was rebooted, and voting started again. After rebooting the experiment machine, the malware attack was ineffective. No votes were flipped after the reboot.   There were no signs of the malware in the experiment machine. This experiment demonstrates that malware cannot persist in a STVM; hence, making malware attacks only effective if the voting machine is never rebooted and/or the USB disk remains connected to the machine, which requires the attack to occur during the election and voters not to notice their votes being flipped and the election officials not notice the inserted USB disk. Both requirements are extremely unlikely with a STVM.

\section{Hand-Marked Paper Ballots vs. Stateless Transparent Voting Machine}
\begin{figure}
    \begin{center}
        \includegraphics [scale=0.40]{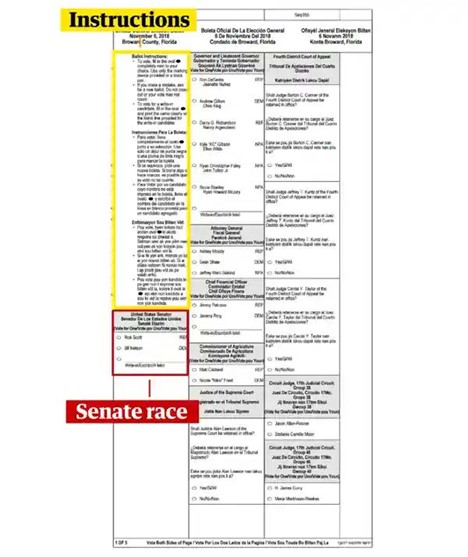}
        \caption{Illustration of Florida 2018 Senate Race Ballot \cite{bad_ballot}}
        \label{badballot}
    \end{center}
\end{figure}

Hand-marked paper ballots (HMPB) are much older than BMD. HMPB are not accessible. Blind voters cannot privately and independently hand-mark a paper ballot. Concerning accessibility, HMPB fail and are not an option for disabled voters. HMPB can be usable, but there have been cases where this was not true. For example, the ballot in the 2018 Florida Senate Race \ref{badballot}, was poorly designed. This resulted in significant undervotes for a highly contested Senate race. Voters missed the contest because it was where instructions normally are located at the bottom of the page. HMPB can be designed to be quite voter-friendly, but this comes with time and effort.  

With respect to security, HMPB have a distinct advantage. Voters must physically place the mark on the ballot themselves. Therefore, it is often assumed that HMPB are voter-verified by default. However, as previously mentioned, voters do make mistakes on HMPB and overlook things; therefore, they can fall short on voter verification. Furthermore, HMPB can suffer from stray marks or improperly filled ovals \cite{ubel2011partisan}. Voters can overvote or undervote. This is only detected if the vote tabulated is set to detect these marks. HMPB can be manipulated for overvote and undervote hacks.

With HMPB, an undervote occurs when no marks are made on the ballot in a specific contest. This is a significant vulnerability from a security perspective that is rarely discussed. If a voter fails to mark any candidates in a specific contest, an insider could make a mark in that contest on the voter's ballot; hence, this is an undervote hack. This hack takes less than three seconds to complete and only requires access and an ink pen. This is impossible to detect unless you catch the insider in the act; therefore, you cannot correct this after the fact. Some optical scanners will look for undervotes and return the ballot if it has undervotes. Despite this, the voter has the right to refuse to vote for any of the candidates in a contest. As a result, undervote ballots are considered legitimate options and should be permissible. 
 
Overvotes occur when the voters mark more than the acceptable number of candidates for a contest. We know HMPB cannot prohibit overvotes. Some implementations rely on the optical scanner to catch overvotes and return the ballot to the voter. Similar to the undervote hack, there is also an overvote hack with HMPB. If a voter selects John Doe for Governor and an insider wants David Rayne to win, the insider can simply mark David Rayne too. This will cause an overvote situation for the voter's ballot for Governor. Like the undervote hack, this is impossible to detect unless you catch the insider in the act, and it's impossible to correct. These attacks do not exist with BMD or STVM, as the system prevents overvotes and may print ``no-selection" for the skipped contest. 

When comparing the HMPB to a STVM, it is important to note that the STVM has accessibility features while the HMPB does not. HMPB are secure against malware and other software hacks; however, they are vulnerable to overvote and undervote hacks, but these do not scale.

\section{Conclusion}
This paper introduced an innovative stateless transparent voting machine (STVM). The STVM provides usability, accessibility, and security for all voters, addressing challenges around voting security presented by ballot-marking devices (BMD) and hand-marked paper ballots (HMPB). This paper discussed the accessibility, usability, and security features compared to BMD and HMPB. The advantages and disadvantages of each voting method were discussed. We also focused on specific attack scenarios attempting to alter the outcome of an election by flipping votes. An experiment with malware that flipped votes was conducted using the STVM and a traditional BMD machine. The experiment demonstrates that, unlike modern BMDs, the STVM is not susceptible to persistent malware attacks that have been discussed in previous papers \cite{feldman2006security,aviv2008security,blaze2017defcon}. The STVM is likely the most secure BMD created at this time. The authors have a working prototype of a STVM that is not designed for production use. A production-level design is still needed. In such a design, the authors propose that the physical casing for the device also be transparent. This will allow anyone to look directly into the internals of the machine and identify any foreign components that have been inserted. AI and computer vision technologies could also be used to efficiently identify foreign components. The STVM prints a voter-verified paper ballot that can also be used in audits, specifically risk-limiting audits (RLA) \cite{stark2009risk}. Practitioners are advised to investigate this style of voting machine further and help establish standard practices and policies around it. The STVM is an innovation that addresses existing challenges in BMD and HMPB. Hopefully, this design can make its way into production and future elections. We hope this design can make its way into production and future elections.

%{\footnotesize \bibliographystyle{acm}
%\bibliography{ref}}
\bibliographystyle{splncs04}
\bibliography{ref}

\begin{thebibliography}{10}
\providecommand{\url}[1]{\texttt{#1}}
\providecommand{\urlprefix}{URL }
\providecommand{\doi}[1]{https://doi.org/#1}

\bibitem{exeter_new_hampshire}
One4all accessible voting system. \url{https://www.exeternh.gov/townclerk/one4all-accessible-voting-system}

\bibitem{verified_voting}
Voting equipment database – prime iii. \url{https://verifiedvoting.org/election-system/prime-iii/}

\bibitem{bad_ballot}
How bad ballot design can sway the result of an election (Nov 2019), \url{https://www.theguardian.com/us-news/2019/nov/19/bad-ballot-design-2020-democracy-america}

\bibitem{abba2017security}
Abba, A.L., Awad, M., Al-Qudah, Z., Jallad, A.H.: Security analysis of current voting systems. In: 2017 International Conference on Electrical and Computing Technologies and Applications (ICECTA). pp.~1--6. IEEE (2017)

\bibitem{aviv2008security}
Aviv, A., {\v{C}}erny, P., Clark, S., Cronin, E., Shah, G., Sherr, M., Blaze, M.: Security evaluation of es\&s voting machines and election management system. In: Proceedings of the conference on Electronic voting technology. pp. 1--13 (2008)

\bibitem{ballotpedia}
Ballotpedia: Voting methods and equipment by state. \url{https://ballotpedia.org/Voting_methods_and_equipment_by_state}

\bibitem{bernhard2020can}
Bernhard, M., McDonald, A., Meng, H., Hwa, J., Bajaj, N., Chang, K., Halderman, J.A.: Can voters detect malicious manipulation of ballot marking devices? In: 2020 IEEE Symposium on Security and Privacy (SP). pp. 679--694. IEEE (2020)

\bibitem{blaze2017defcon}
Blaze, M., Braun, J., Advisors, C.G.: Defcon 25 voting machine hacking village. Proceedings of DEFCON, Washington DC pp. 1--18 (2017)

\bibitem{brooks_2018}
Brooks, D.: New and improved ballot system being used for blind voters in n.h. this election. \url{https://www.concordmonitor.com/blind-voting-election-system-training-nh-19989563} (Sep 2018)

\bibitem{feldman2006security}
Feldman, A.J., Halderman, J.A., Felten, E.W.: Security analysis of the diebold accuvote-ts voting machine  (2006)

\bibitem{gilbert2021study}
Gilbert, J.E., Laurenceau, I., Louis, J.: A study of ballot anomaly detection with a transparent voting machine. Interactions  \textbf{28}(6),  56--61 (2021)

\bibitem{hart-intercivic-nass-2022}
InterCivic], H.: [testing before the vote:best practices for verifying the integrity of your election]. Tech. rep., Hart InterCivic (2022), \url{https://www.nass.org/sites/default/files/2022-06/white-paper-hart-intercivic-nass-summer22.pdf}, white Paper presented at NASS Summer Conference

\bibitem{kortum2021voter}
Kortum, P., Byrne, M.D., Whitmore, J.: Voter verification of ballot marking device ballots is a two-part question: Can they? mostly, they can. do they? mostly, they don't. Election Law Journal: Rules, Politics, and Policy  \textbf{20}(3),  243--253 (2021)

\bibitem{liu2011behavior}
Liu, W., Ren, P., Liu, K., Duan, H.x.: Behavior-based malware analysis and detection. In: 2011 first international workshop on complexity and data mining. pp. 39--42. IEEE (2011)

\bibitem{rivest2008notion}
Rivest, R.L.: On the notion of ‘software independence’in voting systems. Philosophical Transactions of the Royal Society A: Mathematical, Physical and Engineering Sciences  \textbf{366}(1881),  3759--3767 (2008)

\bibitem{sastry2006designing}
Sastry, N., Kohno, T., Wagner, D.: Designing voting machines for verification. In: USENIX Security Symposium (2006)

\bibitem{smarr2017prime}
Smarr, S.A., Sherman, I.N., Posadas, B., Gilbert, J.E.: Prime iii: Voting for a more accessible future. In: Proceedings of the 19th International ACM SIGACCESS Conference on Computers and Accessibility. pp. 335--336 (2017)

\bibitem{stark2009risk}
Stark, P.B.: Risk-limiting postelection audits: Conservative $ p $-values from common probability inequalities. IEEE Transactions on Information Forensics and Security  \textbf{4}(4),  1005--1014 (2009)

\bibitem{ubel2011partisan}
Ubel, P.A., Zikmund-Fisher, B.J.: Partisan vision biases determination of voter intent. PS: Political Science \& Politics  \textbf{44}(1),  81--84 (2011)

\end{thebibliography}

%\section{Acknowledgments}
\end{document}